\documentclass[preprint,aps ,nofootinbib]{revtex4}
\usepackage{graphicx}
\usepackage{amsmath}
\usepackage{amsfonts}
\usepackage{amssymb}
\usepackage{color}%
\usepackage{dcolumn}
\providecommand{\U}[1]{\protect\rule{.1in}{.1in}}

\newcommand{\f}{\begin{equation}}
\newcommand{\ff}{\end{equation}}
\newcommand{\fa}{\begin{eqnarray}}
\newcommand{\ffa}{\end{eqnarray}}

\begin{document}
\title{ Viscous gravitational aether and the cosmological constant problem}
\author{Xiao-Mei Kuang}
\email{xmeikuang@yahoo.cn}
\author{Yi Ling}
\email{yling@ncu.edu.cn}
\affiliation{ Center for Relativistic
Astrophysics and High Energy Physics, Department of Physics,
Nanchang University, 330031, China}

\begin{abstract}
Recently a notion of gravitational aether is advocated to solve the
cosmological constant problem. Through the modification of the
source of gravity one finds that the effective Newton's constant is
source dependent so as to provide a simple but consistent way to
decouple gravity from the vacuum energy. However, in the original
paper the ratio of the effective Newton's constants for pressureless
dust and radiation has an upper bound which is $0.75$. In this paper
we propose a scheme to loose this bound by introducing a bulk
viscosity for the gravitational aether, and expect this improvement
will provide more space for matching predictions from this
theoretical programm with observational constraints.

\end{abstract}
\maketitle
\section{Introduction}
The cosmological constant problem is one of the most important and
deepest problem in theoretical physics. In the seminal review
\cite{Weinberg:1988cp} Weinberg points out that in order to solve
this problem completely one need understand the following two
puzzles. One is why the observed value of the cosmological constant
is so small comparing with the theoretical value of the vacuum
energy predicted by quantum field theory, and the second one is why
the matter density and the dark energy density now observed have the
same order, which is also dubbed as the coincidence problem.
Recently Smolin elaborates the first puzzle into two questions, with
one being qualitative and the other
quantitative\cite{Smolin:2009ti}. Namely, why is the cosmological
constant not of the order of Planck mass? and why is the observed
value is $10^{-120}M_p^2$ after some cancelation from symmetry
breaking.

Recently Afshordi proposed an interesting program to solve the first
cosmological constant problem by introducing a novel notion of
gravitational aether\cite{Afshordi:2008xu}. Following the
thermodynamical description of general relativity by
Jacobson\cite{Jacobson:1995ab}, he conjectured that the usual source
term in Einstein's equation can be modified as
$T_{ab}-\frac{1}{4}g_{ab}T$. Then to preserve the successful
predictions of general relativity and the conversation of the total
energy momentum tensor, or the Bianchi identity, an additional term
$T'_{ab}$ is needed on the right hand side of the equation,
representing the existence of some sort of gravitational aether. One
remarkable result derived from this modification is that the vacuum
energy can be decoupled at appropriate scale. Previously another
approach to degravitate the vacuum energy has been discussed in
\cite{Dvali:2007kt}. However, when degravitating the vacuum energy
from the source, the framework of gravitational aether at the same
time puts an upper bound on the ratio of two effective Newton's
constant for pressureless dust and radiation, which is
$\frac{G_N}{G_R}\leq 0.75$. In the original paper
\cite{Afshordi:2008xu} by Afshordi, it is argued that this ratio
would be preferred by the current observation data, for instance
from the combination of $Ly-\alpha$ forest and CMB
observation\cite{Seljak:2006bg}. Nevertheless, we think this
constraint is far from robust, and as the author noticed, this is a
controversial result if comparing it with the result from Big Bang
Nucleosynthesis\cite{Barrow:2004}, which is given as
$\frac{G_N}{G_R}=0.97\pm 0.09$\cite{Cyburt:2004yc}. In particular
the latter one is a widely admitted and used quantity by most
theoretical physicists and cosmological community. Furthermore, in
this framework it is possible to match the data only under the
condition that the state parameter of the gravitational aether is
very large, namely $\omega' \gg 1$, indicating it is forced to be a
sort of incompressible fluid. Though under some consideration a
large value of the state parameter is desirable, we view this as
another limitation on the nature of the gravitational aether.
Therefore in this paper we intend to loose the constraint on the
ratio of effective Newton's constants in this framework, and expect
this improvement can provide more space for further tests so as to
keep the consistency of this proposed framework with the
observational data.

Our key idea to improve this framework is to generalize the fluid
characteristic of the gravitational aether. Since it is still
beyond the experimental tests, treating the gravitational aether
as a perfect fluid is only an assumption for simplicity. In this
paper we propose that the gravitational aether may be treated as a
fluid with viscosity. It turns out that this treatment will
greatly change the form of the effective Newton's constant, and
thus provide a mechanism to adjust the ratio of the Newton's
constants for matter and radiation.

We organize our paper as follows. In next section we briefly review
the key ideas and results for an ideal gravitational aether and see
how it can degravitate the vacuum energy from gravity. Then we
present our scheme to treat the aether as a viscous fluid in section
three, focusing on the possibility of adjusting the ratio of the
effective Newton's constants. In section four we demonstrate that
this strategy can be extended to the case with multi-fluids. As a
summary, we discuss some open questions and further possible
improvement in section five.

\section{the gravitational aether as a perfect fluid }
In this section we briefly review the approach proposed in
\cite{Afshordi:2008xu} which is supposed to solve the first
cosmological constant problem. In this framework Einstein equation
in a spacetime with contents of common matter and
 gravitational aether is modified as
\begin{equation}\label{aaa}
\frac{1}{8\pi G'}G_{\mu\nu} =
T_{\mu\nu}-\frac{1}{4}Tg_{\mu\nu}+T'_{\mu\nu},
\end{equation}
where $T_{\mu\nu}$ is the ordinary energy-momentum tensor of matter
and $T$ is its trace, while $T'_{\mu\nu}$ is the energy-momentum
tensor of gravitational aether. Adding such a term is to preserve
the divergenceless of the right hand side of Eq.(\ref{aaa}), which
means

\begin{equation}\label{aab}
\nabla_{\mu}T^{\mu\nu}-\frac{1}{4}g^{\mu\nu}\nabla_{\mu}
T+\nabla_{\mu} T'^{\mu\nu}=0.
\end{equation}
Since for the usual matter we have $\nabla_{\mu}T^{\mu\nu}=0$, the
Bianchi identity (\ref{aab}) leads to
\begin{equation}
\frac{1}{4}g^{\mu\nu}\nabla_{\mu} T=\nabla_{\mu} T'^{\mu\nu},
\end{equation}
 which is a {\it dynamical} equation controlling the relations
 between the matter and the gravitational aether.
 Consider the homogeneity of spacetime we may have
\begin{equation}\label{aac}
\frac{1}{4} g^{0 0}\partial_{t} T = \partial_{\mu} T'^{\mu
  0}+\Gamma^{\mu}_{\rho\mu} T'^{\rho 0} + \Gamma^{0}_{\rho\mu}
  T'^{\mu\rho}.
\end{equation}

In the flat FRW metric with the signature $(- + + + )$, we
consider the matter ingredient composed of an ideal fluid with a
constant state parameter $\omega$, thus $T=-(1-3\omega)\rho$. If
one assumes that the gravitational aether is also a simple perfect
fluid with an energy-momentum tensor as
\begin{equation}\label{aad}
T'_{\mu\nu}=p'[(1+\omega'^{-1})u_{\mu} u_{\nu}+g_{\mu\nu}],
\end{equation}
then from Eq.(\ref{aac}) we can derive
\begin{equation}\label{aai}
\frac{dp'}{dt}+3(1+\omega')H
p'=-\frac{3\omega'}{4}(1-3\omega)(1+\omega)H\rho,
\end{equation}
where we have used $\rho=\rho_{0}a^{-3(1+\omega)}$ and
$H=\frac{\dot{a}}{a}$. Equation (\ref{aai}) plays a key role to
degravitate the vacuum energy from gravity.  To understand this we
write down the cosmological equation from the modified Einstein
equation (\ref{aaa}),
\begin{equation}
H^2+\frac{k}{a^2}=\frac{8\pi G_{eff}}{3}\rho,
\end{equation}
with an effective Newton's constant
\begin{equation}\label{aaj}
G_{eff}=(1+\frac{T}{4\rho}+\frac{\rho'}{\rho})G'.
\end{equation}
Substituting the solution to Eq. (\ref{aai}) into above formula one
finds that the effective Newton's constant becomes
\begin{equation}\label{aan}
G_{eff}=
\frac{3}{4}(1+\omega)(\frac{\omega'-\frac{1}{3}}{\omega'-\omega})G',
\end{equation}
where it is required that $\omega'>\omega$ and $\omega'>
\frac{1}{3}$ in order to obtain the inhomogeneous
solution\cite{Afshordi:2008xu}. First of all, we notice that the
effective Newton's constant depends only on the state parameters
of the gravitational aether and matter. In particular the energy
density of aether $\rho'$ plays no role in the final version of
the modified cosmological equation. One remarkable result which
can be read from Eq.(\ref{aan}) is that the effective Newton's
constant would approach to zero whenever $\omega\rightarrow -1$,
indicating that the zero-point energy is decoupled from spacetime
geometry, known as degravitation. However, the price this
mechanism has to pay is that the ratio of the effective Newton's
constants are bounded for various contents of matter.
Specifically, for the cases of pressureless dust era and radiation
era, one finds that
\begin{equation}\label{aaq}
\frac{G_{N}}{G_{R}}=\frac{G_{eff}|_{\omega=0}}{G_{eff}|_{\omega=\frac{1}{3}}}=\frac{3(\omega'-\frac{1}{3})}
{4\omega'}=\frac{3}{4}-\frac{1}{4\omega'},
\end{equation}
which means the ratio of these two effective Newton's constant has
a maximal value which is $0.75$, as the state parameter $\omega'$
goes to infinity. This is a very strong constraint, although in
some circumstances this upper bound still falls in a region
allowed by recent observations and data analysis.

Is there any plausible mechanism to improve this situation? This
is what we intend to do in next section.

\section{Viscous gravitational aether  }

Our main idea is to generalize the fluid characteristic of
gravitational aether, which is treated as a perfect fluid in
previous section. However this is just an assumption. If the
aether has a bulk viscosity, then its energy-momentum tensor is
modified as
\begin{equation}
 T'_{\mu\nu}= [(\tilde{p'}+ \rho')
u_{\mu} u_{\nu} + \tilde{p'} g_{\mu\nu}],
\end{equation}
where the relation between the pressure and the energy density of
the fluid is generalized to
\begin{equation}
\tilde{p'}=\omega' \rho' + \zeta\theta,
\end{equation}
where $\theta$ is the scalar expansion which is proportional to
the Hubble parameter in the context of viscous cosmology, usually
set as $\theta=3H$ in comoving
coordinates\cite{I.Brevik04,Brevik05bj,Odintsov:2006}. $\zeta$ is
the bulk viscosity which in general is a function of energy
densities of multi-fluids as well as their state parameters, and
its specific forms can be found in literature, for instance
\cite{Brevik05bj}. Here we take one of them, setting $\zeta=\alpha
H$ where $\alpha$ may be a function of state parameters $\omega$
and $\omega'$. Then, with the use of the modified Friedmann
equation we can rewrite the pressure as $\tilde{p'}=p'+3\alpha
H^{2}\equiv p'+\tau\rho$, where $\tau$ is independent of energy
densities, but only a function of $\omega$ and $\omega'$ which
should be chosen. In this paper we consider one simple form with
$\tau=\beta \omega' \omega(1 + \omega)$ for the viscosity of the
gravitational aether, where $\beta$ is just a dimensionless
constant. As a result, the pressure has the following form

\begin{equation}
\tilde{p'} = p' + \beta \omega' \omega
 (1 + \omega)\rho.
\end{equation}

Now, with the generalized energy-momentum tensor we turn to
Eq.(\ref{aai}) and find it is changed into
\begin{equation}\label{ac}
\frac{d\rho'}{dt}+3H[\rho'+\omega'\rho'+\beta\omega'\omega
(p+\rho)]=\frac{1}{4}(1-3\omega)\frac{d\rho}{dt}.
\end{equation}

When both state parameters $\omega$ and $\omega'$ are constants,
we obtain a solution to this equation as
\begin{equation}\label{ai}
\frac{\rho'}{\rho}=(\frac{1}{4}-\frac{3}{4}\omega+\beta\omega'\omega)\frac{1+\omega}{\omega-\omega'}.
\end{equation}
We notice that the ratio between the energy densities of matter
and aether keep constant. In contrast to the case of perfect
aether, we find the ratio is not vanishing as $\omega'\rightarrow
\infty$, but approaching to $-\beta \omega(1+\omega)$\footnote{It
means that the energy density of the gravitational aether is
negative whenever the state parameter of matter  $\omega >0$ and
$\beta >0$. Similar situation occurs for an ideal aether fluid,
where the energy density of the aether has to be negative for
$0<\omega <\frac{1}{3}$ and $\omega'>\omega$.}. Since the
modification of energy-momentum tensor will not affect the
relation between $G_{eff}$ and $G'$, substituting Eq.(\ref{ai})
into Eq.(\ref{aaj}) we obtain the effective Newton's constant as
\begin{equation}\label{am}
       G_{eff}
       =\frac{1+\omega}{\omega'-\omega}\left[ \frac{3}{4}(\omega'-\frac{1}{3})-\beta\omega'\omega\right]G'.
\end{equation}
This is the main result in our paper, and several remarks are
given as follows.
\begin{enumerate}
\item First of all, viscous gravitational aether does provide a
similar mechanism to degravitate the vacuum energy from gravity,
as $\omega \rightarrow -1, G_{eff}\rightarrow 0$.

\item In this case the ratio of the effective Newton's constants
for pressureless dust and radiation changes to the form
\begin{equation}\label{ap}
R=\frac{G_{N}}{G_{R}}=\frac{3(\omega'-\frac{1}{3})}{4\omega'}\left[
1+\frac{\beta}{3}\frac{\omega'}{\frac{3}{4}(\omega'-\frac{1}{3})-\frac{1}{3}\beta\omega'}\right].
\end{equation}
Obviously when $\beta\rightarrow 0$,we go back to the case of
ideal aether, i.e. $ R=\frac{3}{4}-\frac{1}{4\omega'}\leq
\frac{3}{4}$. However, when $\beta\neq 0$, we find the ratio can
be larger than $\frac{3}{4}$. For instance, let
$\omega'\rightarrow \infty$, we have
\begin{equation}
 G_{N}=G_{eff}|_{\omega=0}=\frac{3G'}{4},\ \ \ \
 G_{R}=G_{eff}|_{\omega=\frac{1}{3}}=(1-\frac{4\beta}{9})G',
\end{equation}
and
\begin{equation}
R=\frac{G_{N}}{G_{R}}=\frac{3}{4}\frac{1}{1-{4\beta\over 9}}.
\end{equation}
Therefore the ratio can be larger than $\frac{3}{4}$ whenever
$\frac{9}{4}>\beta >0$. In particular, when $\beta= \frac{9}{16}$,
the ratio is exactly equal to one, $R=1$. In Figure one we
illustrate the variation of the ratio with the state parameter of
the gravitational aether when $\beta$ takes various values.

\item In above discussion we intend to lift the upper bound with
an assumption that the aether is a kind of nearly incompressible
fluid, namely $\omega'\gg 1$. As a matter of fact, we may loose
this constraint when the gravitational aether has a viscosity. We
can choose a team of proper values for $\beta$ and $\omega'$ to
let $R=1$. For instance, if we take $\omega'=\frac{2}{3}>\omega$,
we find the ratio $R\rightarrow 1$ as $\beta \rightarrow
\frac{45}{64}$, and the effective Newton's constants are
$G_{N}=G_{R}=\frac{3}{8}G'$. Alternatively, if $\omega'=1>\omega$,
then $R\rightarrow 1$ as $\beta \rightarrow \frac{3}{4}$, and the
effective Newton's constants are $G_{N}=G_{R}=\frac{1}{2}G'$. In
Figure two we illustrate the relation between $1/\omega'$ and
$\beta$ when setting $R=1$, which is nothing but
\begin{equation}
\beta=\frac{9}{16}(1+\frac{1}{\omega'})(1-\frac{1}{3\omega'}).
\end{equation}

\end{enumerate}

\begin{figure}
\center{
\includegraphics[scale=0.75]{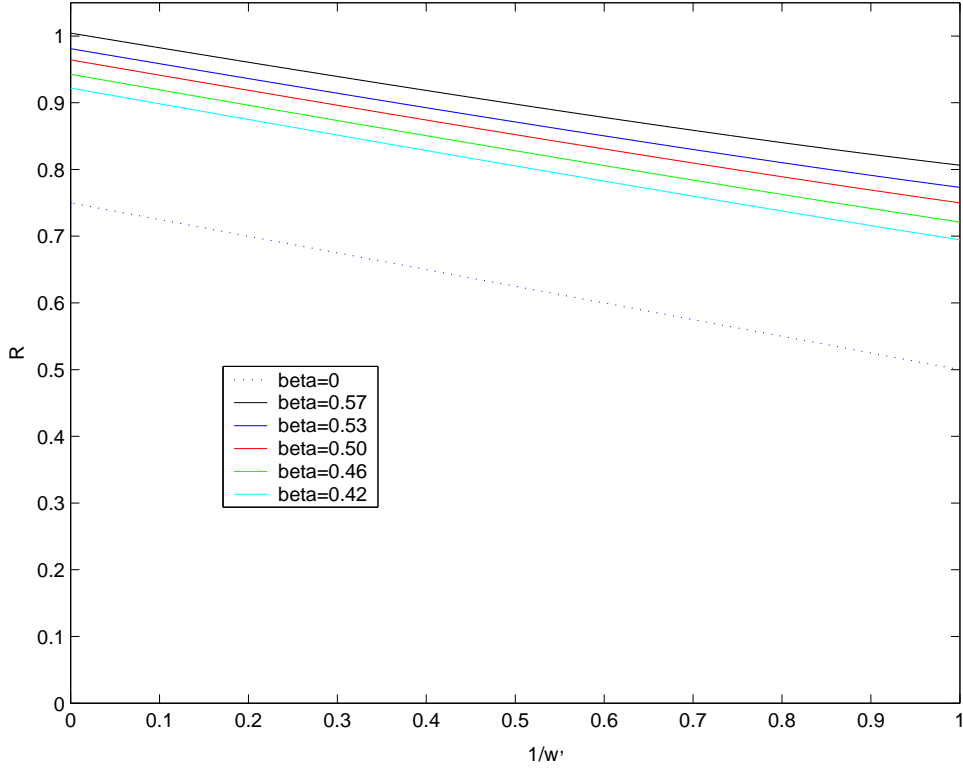}
\caption{The ratio of the effective Newton's constants for
pressureless dust and radiation.}}
\end{figure}

\begin{figure}
\center{
\includegraphics[scale=0.75]{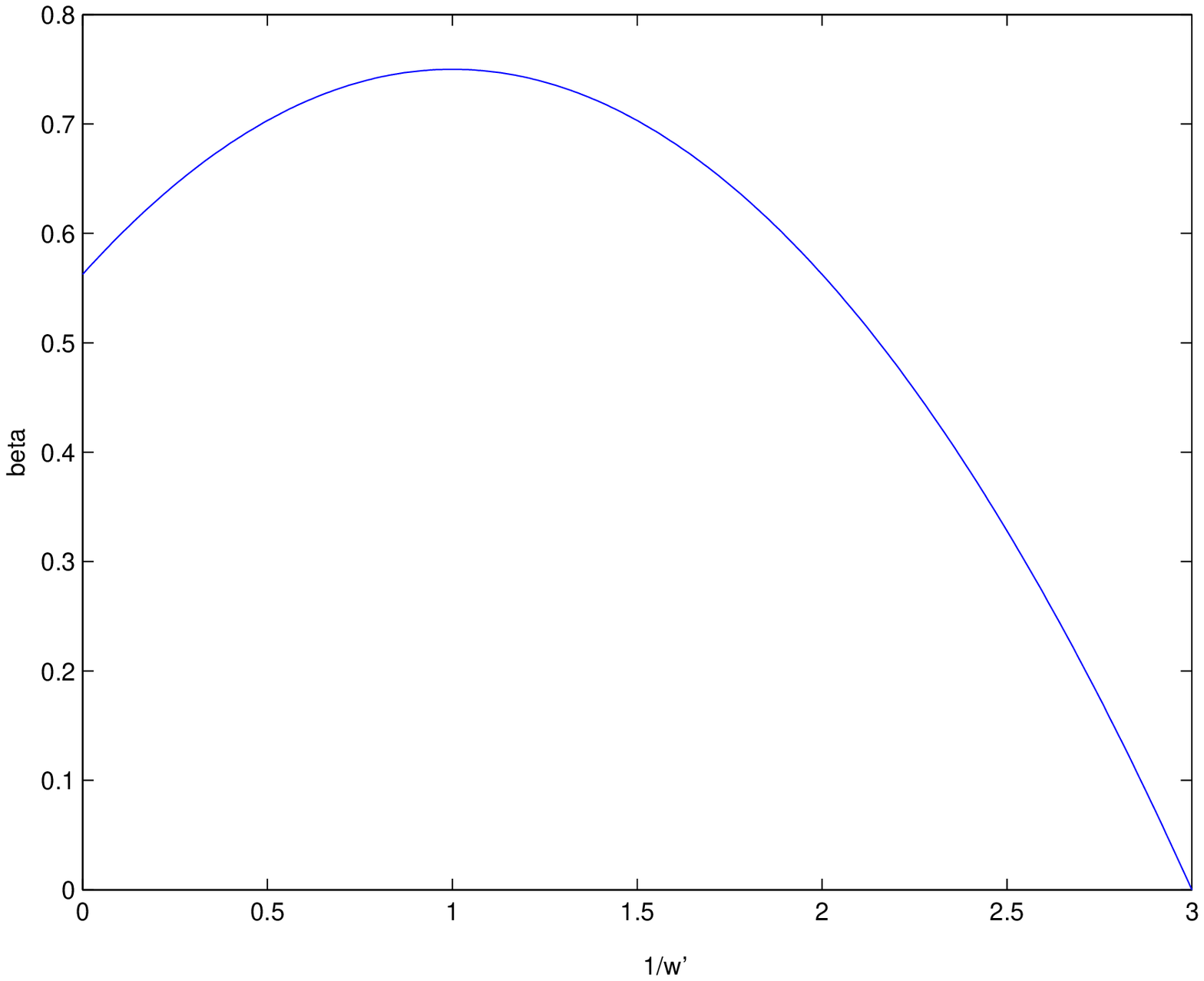}
\caption{The relation between $\beta$ and $\frac{1}{\omega'}$ when
setting $R=1$. }}
\end{figure}

\section{Gravitational aether with multi-fluids }
In previous sections we discussed the effective Newton's constants
could be different for perfect fluids with different state
parameters under the condition that the universe is filled with
the gravitational aether and a single perfect fluid which is
supposed to be dominating in different era of the universe. To
make this proposal more practical we need consider the case that
the gravitational aether is mixed with multi-fluids including all
the ingredients of the universe. Next we demonstrate that the
analysis presented in previous sections can be extended to this
case indeed. For simplicity we consider the gravitational aether
is mixed with two perfect fluids, but all the derivations and
results can be straightforwardly generalized to the case with
multi-fluids.

Consider two perfect fluids with energy densities and constant
state parameters $(\rho_1,\omega_1)$ and $(\rho_2,\omega_2)$,
respectively, we obtain the modified Friedmann equation as

\begin{equation}
H^2+\frac{k}{a^2}=\frac{8\pi G^{eff}_1}{3}\rho_1+\frac{8\pi
G^{eff}_{2}}{3}\rho_2,
\end{equation}
with two effective Newton's constants which have the form
\begin{equation}\label{gef}
G^{eff}_1\rho_1+G^{eff}_2\rho_2=[(\rho_1+\frac{T_1}{4})+(\rho_2+\frac{T_2}{4})+\rho']G'.
\end{equation}

At first we assume that the gravitational aether is a perfect
fluid without viscosity. Then from Eq.(\ref{aac}) we can derive
the equation which constrains the evolution of the gravitational
aether and the fluids as
\begin{equation}\label{acm}
\frac{d\rho'}{dt}+3H(1+\omega')\rho'=\frac{1}{4}(1-3\omega_1)\frac{d\rho_1}{dt}+\frac{1}{4}(1-3\omega_2)\frac{d\rho_2}{dt}.
\end{equation}
To obtain a solution to this equation we firstly notice that from
the conservation equations of fluids
$\dot{\rho_1}+3H(1+\omega_1)\rho_1=0$ and
$\dot{\rho_2}+3H(1+\omega_2)\rho_2=0$, a relation between the
energy densities of two fluids can be obtained
\begin{equation}
\rho_2=\rho_1^{\frac{1+\omega_2}{1+\omega_1}}.
\end{equation}
Substituting this into equation (\ref{acm}), we can solve for
$\rho'$ as a function of the energy density $\rho_1$
\begin{equation}
\rho'=C\rho_1+D\rho_1^m,\label{rhom}
\end{equation}
with
\begin{eqnarray}
m &=& \frac{1+\omega_2}{1+\omega_1}\nonumber\\
C &=& \frac{1-3\omega_1}{4}\frac{1+\omega_1}{\omega_1-\omega'}\nonumber\\
D &=& \frac{1-3\omega_2}{4}\frac{1+\omega_2}{\omega_2-\omega'}.
\end{eqnarray}
Now it is straightforward to obtain the effective Newton's
constants for these two fluids by substituting Eq.(\ref{rhom})
into Eq.(\ref{gef})
\begin{eqnarray}
G^{eff}_1 &=&
\frac{3}{4}(1+\omega_1)(\frac{\omega'-\frac{1}{3}}{\omega'-\omega_1})G'\nonumber\\
G^{eff}_2 &=&
\frac{3}{4}(1+\omega_2)(\frac{\omega'-\frac{1}{3}}{\omega'-\omega_2})G'.
\end{eqnarray}
As a result, we show that the gravitational aether proposal can be
generalized to the case with multi-fluids. In particular, the
effective Newton's constants have the same form as that for a
single perfect fluid.

When the viscosity of the gravitational aether is taken into
account, we propose that the pressure of the aether is modified as
\begin{equation}
\tilde{p'} = p' + \beta \omega' [\omega_1
 (1 + \omega_1)\rho_1+\omega_2
 (1 + \omega_2)\rho_2].
\end{equation}
Then with the same algebra we obtain the effective Newton's
constants for these two fluids as
\begin{eqnarray}\label{amm}
       G^{eff}_1
       &=& \frac{1+\omega_1}{\omega'-\omega_1}\left[ \frac{3}{4}(\omega'-\frac{1}{3})-\beta\omega'\omega_1\right]G'\nonumber\\
       G^{eff}_2
       &=& \frac{1+\omega_2}{\omega'-\omega_2}\left[ \frac{3}{4}(\omega'-\frac{1}{3})-\beta\omega'\omega_2\right]G'.
\end{eqnarray}
They have the same form as that for a single fluid. Therefore we
conclude that our strategy is general and can be applicable to the
case with multi-fluids.
\section{Discussion and Conclusions}

The notion of gravitational aether is introduced to solve the first
cosmological constant problem. In this mechanism the effective
Newton's constant is matter dependent such that the vacuum energy
can be decoupled from gravity, but the price is that the ratio of
effective Newton's constants for pressureless dust and radiation is
bounded, no larger than $0.75$. In this paper we proposed that this
difficulty may be overcome by generalizing the ideal gravitational
aether to the one with viscosity. Moreover, the aether need not be a
kind of incompressible fluid.

To obtain such effects we have taken a special form of the viscous
term. In particular, it depends on the energy density of ordinary
matter rather than that of the gravitational aether, which looks
peculiar. It means that the viscosity is not an intrinsic property
of the gravitational aether itself, but generated whenever the
aether is mixed with matter fluids. Although we may change its
form into one proportional to the square of the Hubble parameter
through the Friedmann equation, we think this form of viscosity
calls for further understanding.

To push the proposal of gravitational aether forward, we need do
more, specially in the interpretation of current accelerating
expansion of the universe and the proportion of the dark energy in
contents of cosmological ingredients\cite{Afshordi:2009}. Since
our strategy can be extended to the case with multi-fluids, the
program along this direction is under progress and will be
presented elsewhere.

\begin{acknowledgments}
We are grateful to Niayesh Afshordi and Hongbao Zhang for reply and
useful discussion. This work is partly supported by
NSFC(Nos.10663001,10875057), JiangXi SF(Nos. 0612036, 0612038), Fok
Ying Tung Education Foundation(No. 111008) and the key project of
Chinese Ministry of Education(No.208072). We also acknowledge the
support by the Program for Innovative Research Team of Nanchang
University.

\end{acknowledgments}

\end{document}